\documentclass[journal=apchd5,manuscript=article]{achemso}

\usepackage[version=3]{mhchem} 
\usepackage{siunitx}

\newcommand{\micras}[0]{\si{\micro\metre}}
\newcommand{\matriz}[1]{\mathbf{#1}}

\author{X. Garcia-Santiago}
\affiliation[INT]
{Institute of Nanotechnology, Karlsruhe Institute of Technology, Karlsruhe, Germany}
\email{xavier.garcia-santiago@kit.edu}
\author{M. Hammerschmidt}
\affiliation[JCM]
{JCMwave GmbH, Berlin, Germany}
\alsoaffiliation[ZIB]
{Computational Nano Optics, Zuse Institute Berlin, Berlin, Germany}
\author{J. Sachs}
\affiliation[MPIIS]
{Max Planck Institute for Intelligent Systems, Stuttgart, Germany}
\author{S. Burger}
\affiliation[ZIB]
{Computational Nano Optics, Zuse Institute Berlin, Berlin, Germany}
\alsoaffiliation[JCM]
{JCMwave GmbH, Berlin, Germany}
\author{H. Kwon}
\affiliation[MPIIS]
{Max Planck Institute for Intelligent Systems, Stuttgart, Germany}
\author{M. Kn{\"o}ller}
\affiliation[IANM]
{Institute for Applied and Numerical Mathematics, Karlsruhe Institute of Technology, Karlsruhe, Germany}
\author{T. Arens}
\affiliation[IANM]
{Institute for Applied and Numerical Mathematics, Karlsruhe Institute of Technology, Karlsruhe, Germany}
\author{P. Fischer}
\affiliation[MPIIS]
{Max Planck Institute for Intelligent Systems, Stuttgart, Germany}
\alsoaffiliation[IPC]
{Institute of Physical Chemistry, University of Stuttgart, Stuttgart, Germany}
\author{I. Fernandez-Corbaton}
\affiliation[INT]
{Institute of Nanotechnology, Karlsruhe Institute of Technology, Karlsruhe, Germany}
\author{C. Rockstuhl}
\affiliation[INT]
{Institute of Nanotechnology, Karlsruhe Institute of Technology, Karlsruhe, Germany}
\alsoaffiliation[TFP]
{Institute of Theoretical Solid State Physics, Karlsruhe Institute of Technology, Karlsruhe, Germany}
\alsoaffiliation[MPSP]
{Max Planck School of Photonics, Germany}

\title{Towards maximally electromagnetically chiral scatterers at optical frequencies}

\begin{document}






\begin{abstract}
  Designing objects with predefined optical properties is a task of fundamental importance for nanophotonics, and chirality is a prototypical example of such a property, with applications ranging from photochemistry to nonlinear photonics. A measure of electromagnetic chirality with a well-defined upper bound has recently been proposed. Here, we optimize the shape of silver helices at discrete frequencies ranging from the far infrared to the optical band. Gaussian process optimization, taking into account also shape derivative information of the helices scattering response, is used to maximize the electromagnetic chirality. We show that the theoretical designs achieve more than 90 percent of the upper bound of em-chirality for wavelenghts \SI{3}{\micro\meter} or larger, while their performance decreases towards the optical band. We fabricate and characterize helices for operation at \SI{800}{\nano\meter}, and identify some of the imperfections that affect the performance. Our work motivates further research both on the theoretical and fabrication sides to unlock potential applications of objects with large electromagnetic chirality at optical frequencies, such as helicity filtering glasses. We show that, at \SI{3}{\micro\meter}, a thin slab of randomly oriented helices can absorb 99 percent of the light of one helicity while absorbing only 9 percent of the opposite helicity. 
\end{abstract}

\section{Introduction}

Chirality is a property usually judged only through geometrical considerations. The definition for chirality has been put forward by Lord Kelvin\cite{kelvin1894molecular}. It states that an object is chiral if it cannot be superimposed onto its mirror image, called enantiomer, by any means of rotation or translation. In contrast, if the objects can be superimposed on each other, they are considered achiral. 

The concept of chirality is important for many scientific disciplines. In chemistry, for example, many molecules are chiral, and certain reactions are only triggered by one of the two enantiomers. As most physical properties of the two enantiomers are identical, we can only distinguish them by probing their response with another chiral object or a chiral interaction. Light is an excellent tool for this purpose as chiral molecules or any other chiral object interact with electromagnetic fields differently depending on the handedness of the illumination\cite{mun2020electromagnetic}. The helicity of the field expresses the handedness. For a single plane wave, it concerns left- or right-handed circularly polarization.  

The differential interaction of electromagnetic fields with chiral objects is used to design devices that increase the cross-section of a certain molecular reaction\cite{richardson2015dual,yang2017circularly,he2018dissymmetry}, molecular interaction\cite{schaferling2012tailoring,mohammadi2018nanophotonic,hentschel2017chiral,schaferling2014helical,jeong2016dispersion,vestler2018circular,graf2019achiral,feis2020helicity,kelly2005study}, or that can filter polarization\cite{chadha2014all,rangelov2019broadband}. Suppose one aims to design a device that maximizes the difference in the interaction with fields of a specific handedness. In this case, it could be reasonable to think about maximizing the chirality of the object on the ground of geometrical consideration. The first problem with this idea is that it has already been shown that a consistent measure for quantifying the typical geometrical definition of chirality does not exist \cite{fowler2005quantification}. Many proposals to measure the chirality of an object have been put forward, e.g., in \cite{buda1992hausdorff}, but all of them present certain inconsistencies \cite{buda1992onquantifying}. A measure for the electromagnetic chirality, $\chi$, was recently proposed to lift that issue \cite{fernandez2016objects}. This measure assigns a degree of electromagnetic chirality, em-chirality for brevity, to objects based on how differently the object interacts with all the possible illumination fields of different helicity. While the definition of $\chi$ encompasses general illuminations, a single frequency version can be obtained by restricting the illumination to monochromatic fields.

The em-chirality is consistent with the chirality of an object. Any achiral object will have a $\chi$ value of zero at all frequencies. The link between the em-chirality and further electromagnetic quantities, such as circular dichroism, and an interaction-based measure of geometrical chirality were studied in \cite{gutsche2020role}.

One salient property of $\chi$ is that its upper-bound is given by the total interaction cross-section of the object. The latter can be used to normalize $\chi$. The values of the normalized measure, $\overline{\chi}$, range from 0 to 1. The upper bound at 1 gives rise to the notion of maximally em-chiral objects. If an object is maximally em-chiral at some frequency, this object would be invisible to any field of such frequency which is a pure state of one of the two eigenstates of the helicity operator, see Ref.~\mbox{(\!\!\citenum{tung1985group})} Chap.~8.  Thus, this object would be ideal for angle-independent polarization filters, as proposed in\cite{fernandez2016objects}, and it would also be a cornerstone for systems requiring large interaction differences with the two helicities.

The question whether one can obtain these objects, in reality, remained, however, open. Analytical designs to achieve such objects can be approached by assuming perfect electric conductors and assuming that the object is made from arbitrarily thin wires \cite{wheeler1947helical}. Concerning real-world objects, it has been shown that by considering these analytical designs, a helix made from silver could be designed that shows a $\overline{\chi}$ above 0.9, however, only at far infrared wavelength (e.g., at 200 \micras{} \cite{fernandez2016objects}). Up to now, it is not clear if objects with high $\overline{\chi}$ can be obtained at optical frequencies or in the near-infrared, where silver is no longer a perfect conductor. The onset of a kinetic inductance in the metal usually prevents the extrapolation of analytical designs at radio-frequencies towards the visible. 

In this work, we look into whether it is possible to design scatterers with a high em-chirality at frequencies as large as the optical frequencies using suitable techniques from the field of computational material design. We limit the search to silver helices. The reason for that is that helices are canonical chiral objects that have been shown to present strong differences in their interaction with fields of different handedness \cite{wheeler1947helical, gansel2010gold, chadha2014all, chadha2014comparative, gutsche2016locally,martens2021origin,Kuzyk2012}. High values of em-chirality have already been shown for helices. Limiting the search to helices also simplifies the optimization process compared to more general free-form wires. For example, one then needs to apply complicated constraints to avoid overlapping the object with itself due to its finite thickness.

To identify silver helices with high $\overline{\chi}$ values for different illumination wavelengths, we combined numerical techniques to compute the em-chirality of complex geometries and also their shape derivatives with a machine learning algorithm for global optimization. We run different optimizations covering a wavelength range from 150 \micras{} down to 500 nm to explore the frequency dependency of $\overline{\chi}$ for optimal helices as well as their geometrical properties. Promising designs were realized through nanofabrication, and their optical response was experimentally characterized. The optical response of the optimal design and that of the fabricated sample showed discrepancies. Nevertheless, a refined analysis considering both the deviating geometry of the helices, as extracted from SEM images, and material properties accommodating oxidation of silver, clearly explained the difference. The results made it possible to numerically assess the suitability of the designed and fabricated samples in helicity filtering glasses. We demonstrate that, at the operational wavelength of \SI{3}{\micro\meter}, a \SI{8}{\micro\meter} slab of randomly arranged helices at 5 percent filling factor is able to absorb 99 percent of the light of unwanted helicity while absorbing only 9\% of the light of the desired helicity. For the same rejection of the unwanted helicity, both the optimal design at \SI{800}{\nano\meter} and the fabricated helices would need a slab of approximately \SI{20}{\micro\meter} which would absorb 46 percent and 79 percent of the desired light, respectively. Our results indicate that both theoretical and fabrication improvements are needed to extend the performance of infrared helices onto the optical band.

This work is structured in three parts. First, we show the optimization results, and we then analyze the response of different helices obtained during the numerical optimization process. Second, after obtaining the optimal designs, we analyze the response of experimentally realized helices for one of the optimal designs obtained at optical frequencies. Third, we use a simple homogenization theory model to analyze the performance of some of the designs obtained for creating an omni-angle circular polarization filter.

\section{Optimization results}

The normalized em-chirality, $\overline{\chi}$, of an object is defined\cite{fernandez2016objects} based on the T-matrix of the object~(e.g., Ref.~\mbox{(\!\!\citenum{mishchenko2002scattering})} Chap. 5) using a basis of multipoles of pure helicity,

\begin{equation}
\label{eq:chibar}
    \overline{\chi} = \frac{\left\lVert\begin{bmatrix}\mathrm{svd}\left(\matriz{T}^{+,+}\right)\\\mathrm{svd}\left(\matriz{T}^{+,-}\right)\end{bmatrix}-\begin{bmatrix}\mathrm{svd}\left(\matriz{T}^{-,-}\right)\\\mathrm{svd}\left(\matriz{T}^{-,+}\right)\end{bmatrix}\right\rVert}{\lvert\lvert\matriz{T}\rvert\rvert}
\end{equation}
where $\mathrm{svd}\left(\matriz{A}\right)$ denotes a vector containing the singular values of matrix $\matriz{A}$, $\lvert\lvert\matriz{A}\rvert\rvert$ denotes the Frobenius norm of matrix $\matriz{A}$, and $\matriz{T}^{\pm,\pm}$ are the submatrices of the T-matrix for specific helicities of the illumination and scattered fields. 
The T-matrix in the helicity basis can be obtained as a simple linear combination\cite{fruhnert2017computing} of the elements of the T-matrix expanded in the basis of electric and magnetic multipoles.

We use the finite element method solver JCMsuite to numerically calculate the T-matrices of the helices\cite{pomplun2007adaptive,burger2008jcmsuite}. Besides calculating the T-matrix, we exploit the calculation of shape derivatives of the T-matrix based on the forward method\cite{burger2013fast,kepler2010sensitivity,hughes2019forward}. This technique generates, within the same finite element simulation, both the em-chirality of the object and its shape derivatives with respect to the geometrical parameters of the helix. To the best of our knowledge, this is the first time that a method to calculate shape derivatives of T-matrices for generally complex scatterers is presented and used for optimization. The method can be used in the design of scatterers for many applications that rely on a multipole description of the scatterer response\cite{dezert2017isotropic,dezert2019complete,zambrana2013duality,slivina2019insights,mun2020multipole}. The details of the FEM simulations and how to compute the T-matrices and their shape derivatives are presented in section~\ref{sec:Methods}.


Four parameters describe the geometry of the helices: the radius of the helix spine, $R_{\mathrm{h}}$, the thickness of the helix wire, $T_{\mathrm{w}}$, the pitch of the helix $P_{\mathrm{h}}$, and the number of turns, $N_{\mathrm{turns}}$. 
To optimize the different helices, we combine the finite element method to calculate the em-chirality with a Bayesian optimization based on Gaussian processes\cite{pelikan1999boa,snoek2012practical} as the optimization algorithm. This algorithm was shown to be an efficient global optimization method for medium dimensional optimization problems with expensive objective functions\cite{schneider2019benchmarking}. Furthermore, Bayesian optimization can incorporate derivative information of the objective function in a simple and elegant way (e.g., Refs.~\mbox{(\!\!\citenum{williams2006gaussian})} Chap. 9.4, \mbox{(\!\!\citenum{solak2003derivative,wu2017BOgradients})}). The extra information provided by the shape derivatives can improve the optimization convergence rate. We provide the derivatives of the four parameters of the design space to the Gaussian process.

The optimizations are performed for different specific design vacuum wavelengths, $\lambda _{\mathrm{d}}$, ranging from $\lambda _{\mathrm{d}}$ = 150 \micras{} to $\lambda _{\mathrm{d}}$ = 500 nm. Our goal is to see whether we can obtain helices with large $\overline{\chi}$ values along the entire infrared spectrum and also at optical wavelengths. In case this is not possible, we aim to study at which spectral region the material properties of the silver do not allow helices with large em-chirality to be obtained. That breakdown of the functionality is expected to happen towards optical wavelengths. In this spectral region, silver tends to be a poor metal due to the onset of absorption. Also, silver is no longer metallic above the plasma frequency, and it would not sustain localized plasmon polaritons that are key for a resonant light-matter-interaction in metals. The permittivity for silver at each wavelength is interpolated from two databases\cite{mcpeak2015plasmonic,hagemann1975optical} depending on the frequency range. 

We start the optimization at long wavelengths because of the availability of designs that already work quite well \cite{fernandez2016objects}. These known designs allow for a rough estimation of the parameter space within which an optimization is feasible.

In total, we performed 12 optimization runs, each one corresponding to a different design wavelength of interest. Each run took between one and two days, during which we evaluated 500 different points of the design space. Those points are chosen by the Bayesian optimization algorithm. Section~\ref{sec:Methods} presents further details of the optimization process. 

\begin{figure}[h!]
\centering\includegraphics[width=0.8\linewidth]{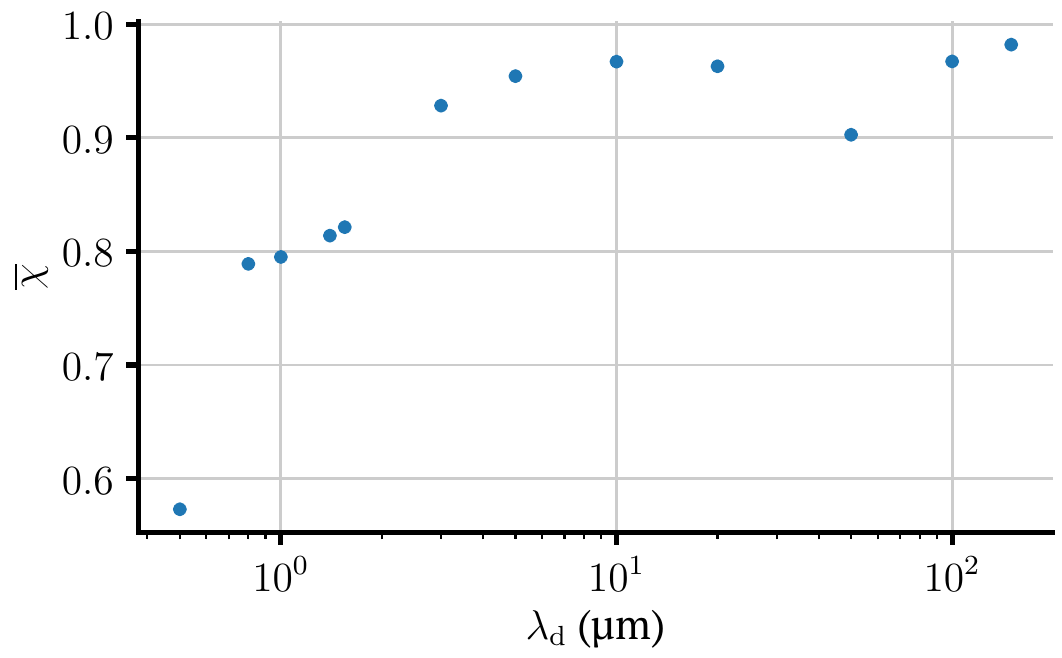}
\caption{Electromagnetic chirality of silver helices with optimized geometrical parameters at different design wavelengths, $\lambda_{\mathrm{d}}$. Each point shows the normalized em-chirality value, $\overline{\chi}$, of the optimal helix found for that specific operational wavelength.}
\label{fig:optimal_chirality_values}
\end{figure}

\begin{figure}[h!]
\centering\includegraphics[width=0.8\linewidth]{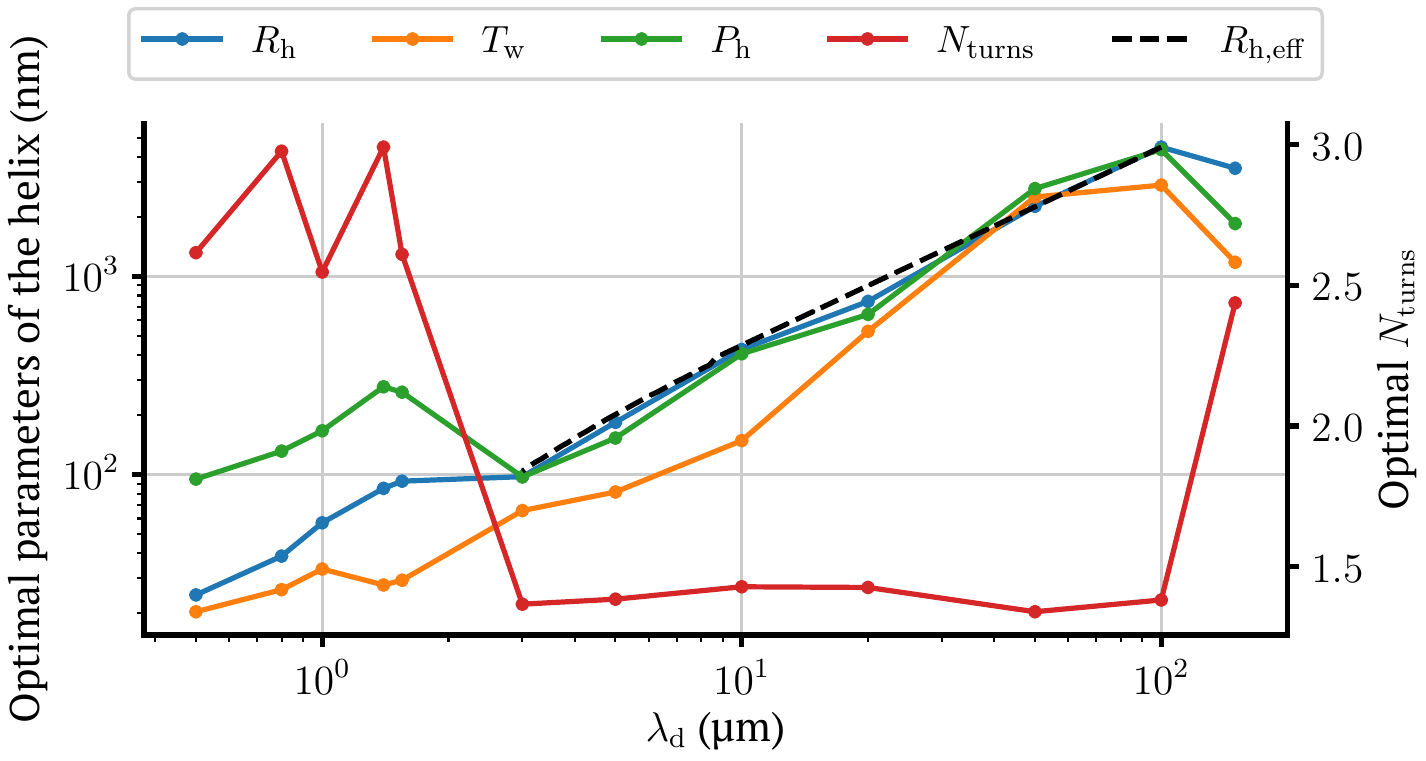}
\caption{Geometrical parameters of each optimal helix obtained at each design wavelength $\lambda _{\mathrm{d}}$. The parameters defining the shape of the helix are the radius of its spine, $R_{\mathrm{h}}$, the thickness of the helix wire, $T_{\mathrm{w}}$, the pitch per turn of the helix, $P_{\mathrm{h}}$, and the total winding number, $N_{\mathrm{turns}}$. $R_{\mathrm{h,eff}}$ shows the downscaling of the value of $R_{\mathrm{h}}$ with the wavelength, $\lambda _{\mathrm{d}}$, calculated with a scaling design rule for plasmonic materials\cite{novotny2007effective}.}
\label{fig:optimal_helices}
\end{figure}

The results of the optimizations are shown in Figs.~\ref{fig:optimal_chirality_values} and~\ref{fig:optimal_helices}. The optimal em-chirality values found for the different wavelengths are shown in Fig.~\ref{fig:optimal_chirality_values}. Figure~\ref{fig:optimal_helices} shows the corresponding parameters of the optimal helices. As we can see in Fig.~\ref{fig:optimal_chirality_values}, we found helices with $\overline{\chi}$ values larger than 0.9 at wavelengths down to~3 \micras{}. Below this wavelength, $\overline{\chi}$ starts to drop, and at 500~nm, we could not find helices with values higher than 0.58. 

Inspecting both Fig.~\ref{fig:optimal_chirality_values} and ~\ref{fig:optimal_helices}, one can distinguish two wavelength regions with different behaviors. The optimal designs found at wavelengths above $\lambda _{\mathrm{d}}$ = 3~\micras{} present a similar shape. In this region, the optimal helices have a similar number of turns, and also the ratio between the radius of the helix spine $R_{\mathrm{h}}$ and the pitch of the helix keeps more or less constant. Moreover, these last two parameters seem to scale linearly with the operational wavelength. All these observations can be explained based on the approximation of the material properties of silver in this wavelength region as a perfect electric conductor. The black dashed line plotted in Fig.~\ref{fig:optimal_helices} represents the downscaling of the optimal helix obtained at $\lambda _{\mathrm{d}}$ = 100~\micras{} following the design scaling rule for plasmonic materials described in\cite{novotny2007effective}. This scaling factor differs slightly from unity only in the region between 3 to 10~\micras{}. Some small deviations from this scaling rule present in the optimization results could be explained by the fact that the FEM simulations take into account the effects of a non-zero thickness of the wire. Even in cases where it is much smaller than the wavelength, this can play an important role in the scattering response\cite{jones1990thin}.

For wavelengths below 3~\micras{}, the plasmonic properties of silver disappear. Thus, it will be increasingly complicated for the helix to sustain a suitable balance between electric and magnetic dipolar resonance necessary to obtain large values for the em-chirality\cite{fernandez2013electromagnetic,fernandez2020computation,hoeflich2019resonant}. The decrease can also be rationally appreciated from considering the limiting behavior. Halving the size of the helix at far-infrared frequencies allows doubling the operational frequency. However, the possibility of silver to sustain localized plasmon polaritons is bound to a lower wavelength of approximately 360~nm due to the underlying material dispersion. Therefore, halving the size does not allow to halve the operational wavelength when approaching the visible part of the spectrum. Instead, the resonance wavelength is adiabatically pushed towards the lower bound. This implies a decreasing ratio of the absolute size of the helix to the operational wavelength. In turn, this suggests that the resonance strength decreases. In particular, the magnetic dipolar response can only be weakly excited for a decreasing size of the object. This effect lowers the achievable magnetic response that is required for large values of em-chirality. Similar effects have been encountered when studying the scaling behavior of, e.g., split-ring resonators\cite{ishikawa2005negative,klein2006single,zhou2005saturation}. 

\begin{figure}[h!]
\centering\includegraphics[width=0.8\linewidth]{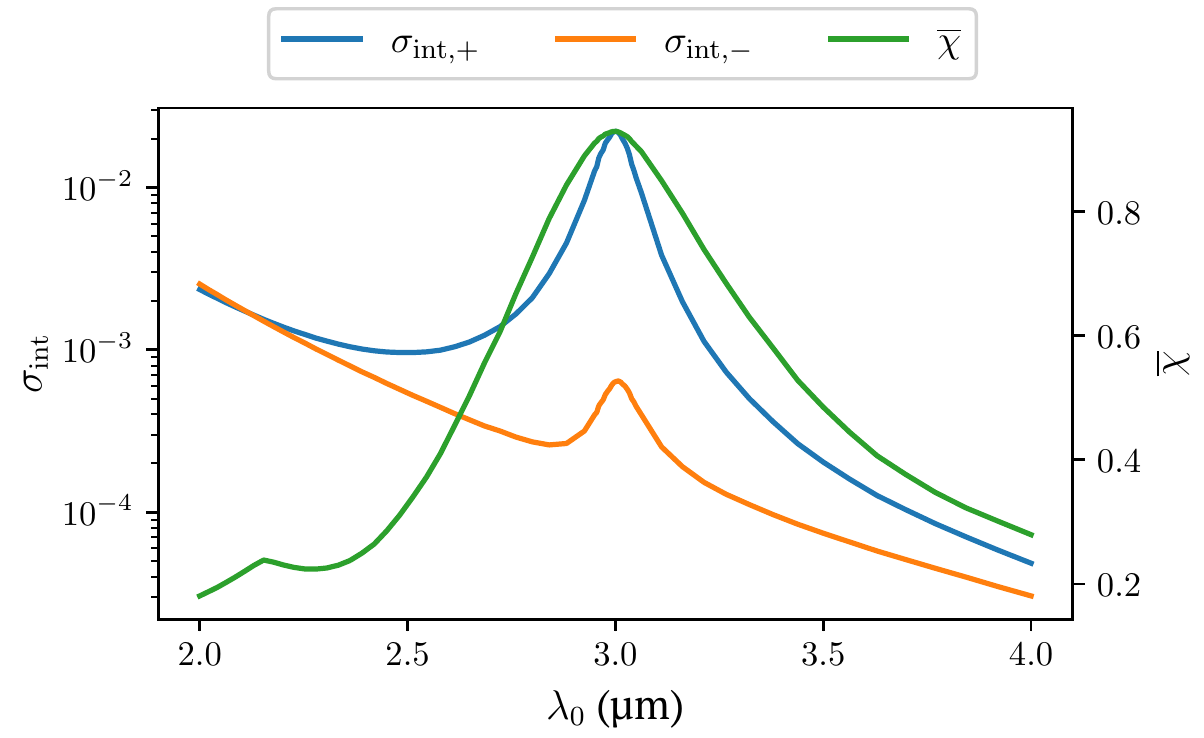}
\caption{Spectral dependency of the em-chirality and the interaction cross-section $\sigma_\mathrm{int}$ for the helix optimized for an operational wavelength of $\lambda _{\mathrm{d}}$ = 3~\micras{} (100~THz). The interaction cross-section is split into the values of the interaction with fields of both helicities.}
\label{fig:frequency_sweep}
\end{figure}

To analyze the properties of the optimal helices, we take as an example the helix optimized for an illumination wavelength of 3~\micras{}. In Fig.~\ref{fig:frequency_sweep}, we plot its em-chirality and interaction cross-section as a function of the wavelength. The interaction cross-section is the sum of the squared singular values of respective T-matrices \cite{fernandez2016objects}. It is a measure of how much an object interacts with light at a certain wavelength, and appears in the denominator of Eq.~(\ref{eq:chibar}). We see that the maximum of the em-chirality corresponds to a peak of the interaction cross-section. This maximum implies that it is not an undesirable optimum where the em-chirality is high and the interaction strength is low, but rather that it is a resonance. Figure~\ref{fig:frequency_sweep} also shows an order of magnitude difference in the interaction of the helix with fields of different helicity. We notice that an object so close to the maximal possible em-chirality barely interacts with light of one helicity while interacting strongly with the opposite helicity. Such a response is in stark contrast to an achiral object that interacts identically with light of both helicities.

\begin{figure}[h!]
    \includegraphics[width=\linewidth]{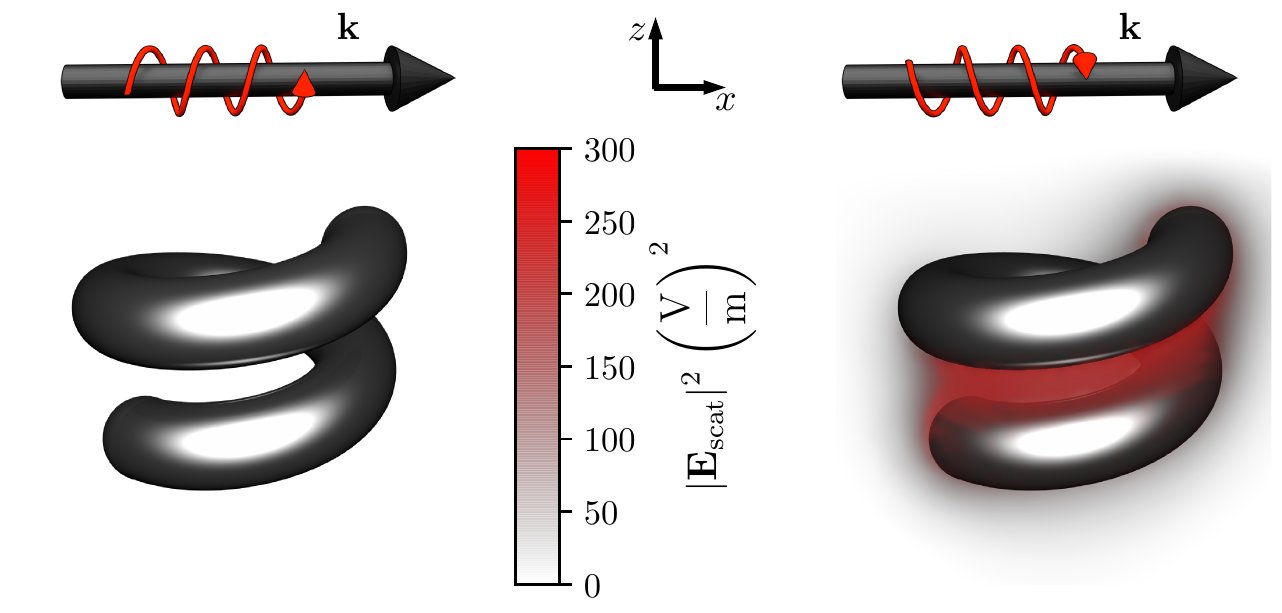}
\caption{Scattered field intensity obtained when the optimized helix for an operational wavelength of 3~\micras{} is illuminated with two plane waves of different circular polarization and the same wavelength and direction. The plane waves have a vacuum wavelength of 3 \micras{}, and their wave vector is parallel to the x-axis. The electric field strength of the plane waves is 1~V/m. The left and right figures show the fields for the left and right circular polarization, respectively.}
\label{fig:helices_scattering_intensity}
\end{figure}

This difference in the interaction with fields of different helicity is also visible in Fig.~\ref{fig:helices_scattering_intensity}. The figure shows the intensities of the scattering fields close to the helix upon illumination with a plane wave propagating in the x-direction. The plane wave is circularly polarized, either left- or right-handed. The intensity of the scattered field is approximately two orders of magnitude higher for the right circularly polarized plane wave illumination, consistent with the results obtained for the interaction cross-section. 

\begin{figure}[h!]
\centering\includegraphics[width=0.95\linewidth]{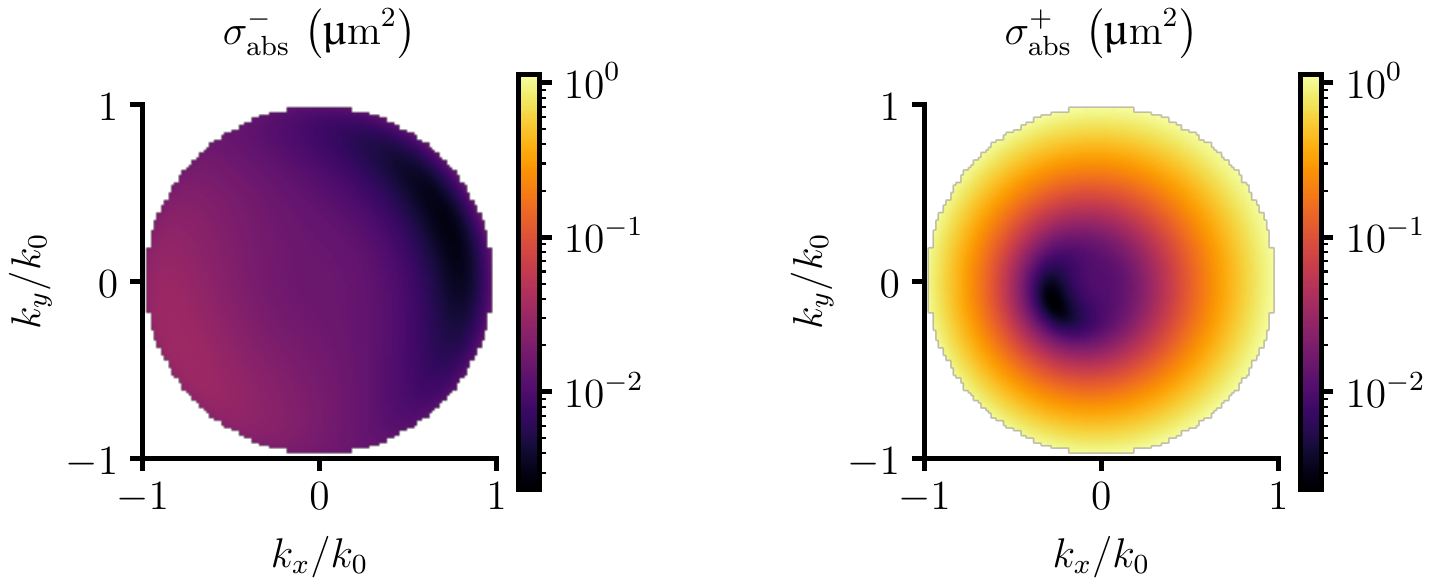}
\caption{Absorption cross-sections, $\sigma_\mathrm{abs}$, of the helix optimized for an operational wavelength of 3~\micras{} when it is illuminated with plane waves of different circular polarization. The plot shows $\sigma _\mathrm{abs}$ as a function of the direction of the plane waves, represented in k-space. The helix winds around the $z$-axis}
\label{fig:sigmas_3_um}
\end{figure} 

Figure~\ref{fig:sigmas_3_um} shows the absorption cross-section for the different directions of illumination plane waves and both circular polarizations. We see how the largest difference between polarizations arises for plane waves perpendicular to the axis of the helix. For this direction, the absorption is two orders of magnitude larger for the right circularly polarized light. The large differences in the interaction and absorption cross-sections provided by this helix for different illumination directions render it a good candidate to be used in a direction-independent circular polarization filter, as proposed in \cite{fernandez2016objects}. 

\section{Experimental results}

\begin{figure}[h!]
\centering\includegraphics[width=0.95\linewidth]{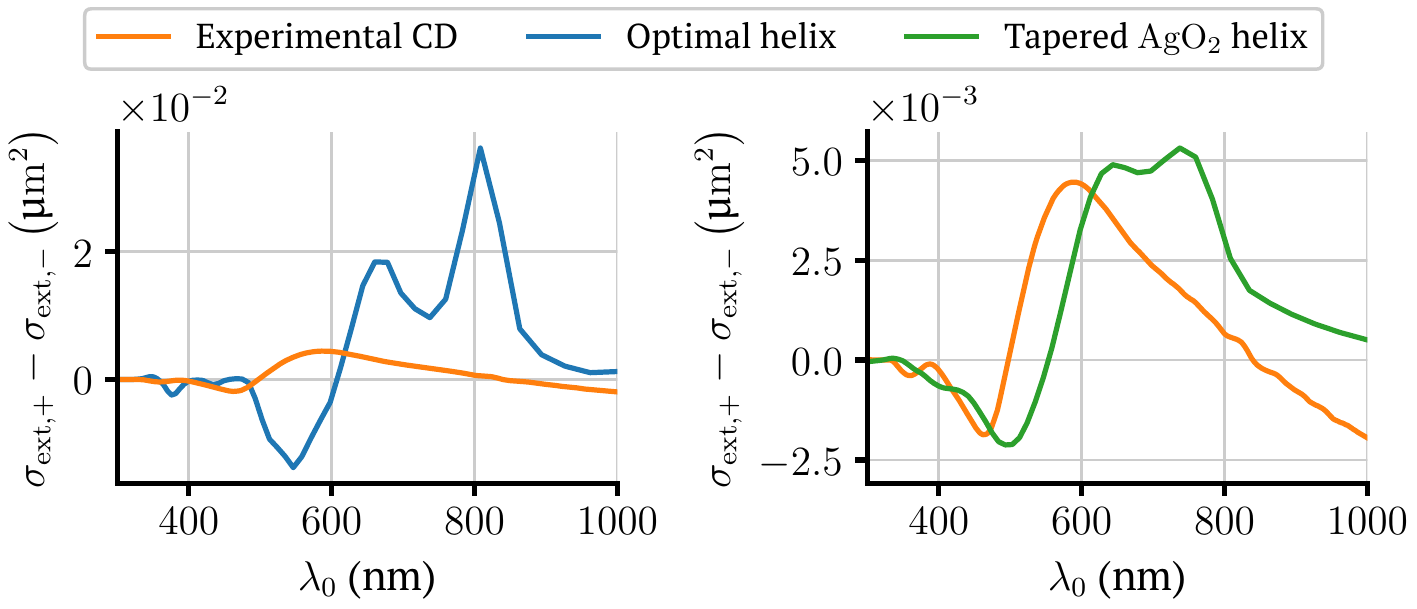}
\caption{\textbf{a}: Experimental rotationally averaged differential extinction cross-sections $\sigma_\mathrm{ext,+}-\sigma_\mathrm{ext,-}$ with respect to the vacuum illumination wavelength, $\lambda _0$, for the fabricated helices (orange) and the numerical values for the optimized helix (blue). The values of the experimental results are obtained from the experimental CD signal, in millidegrees, based on the concentration of helices into the solvent. The details of the conversion are explained in section~\ref{sec:Methods}. \textbf{b}: The orange line shows the same results as in \textbf{a}. The green line shows the numerical results of the differential extinction cross-section for a partially oxidized tapered helix. The shape of the helix is shown in Fig.~\ref{fig:SEM}c.}
\label{fig:experimental_results}
\end{figure}

\begin{figure}[h!]
\centering\includegraphics[width=0.95\linewidth]{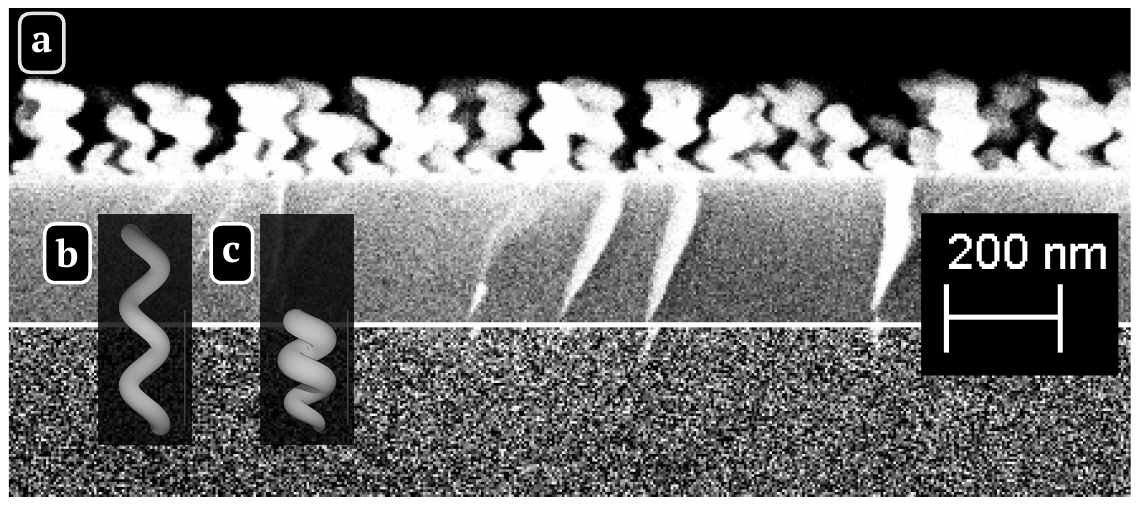}
\caption{\textbf{a}: SEM image of the fabricated helices. The helices were fabricated using a vapor deposition technique described in the Methods section. \textbf{b}: Rendered image of the helix that came as a result of the numerical optimization process. Its parameters are: $R_{\mathrm{h}}$ = 27.4~nm, $T_{\mathrm{h}}$ = 35~nm, $P_{\mathrm{h}}$ = 137~nm, and $N_{\mathrm{turns}}$ = 2.4. Its helicity differential extinction cross-section corresponds to the blue line of Fig.~\ref{fig:experimental_results}a. \textbf{c}: Rendered image of the CAD model of a tapered helix used for trying to reproduce the experimental results numerically. The green line shows its helicity differential extinction cross-section signature in Fig.~\ref{fig:experimental_results}b. All images a, b, and c present the same scale and, therefore, can be directly compared.}
\label{fig:SEM}
\end{figure}

In a further step, we fabricated the optimal helix obtained for the design wavelength of 800 nm and measured its optical response. The helices are fabricated with a vapor deposition technique described in section~\ref{sec:Methods}. They are made with 95 percent silver and a 5 percent of Cu in their composition. 
Because the CD measurement is performed with helices immersed into Milli-Q water, the helix geometry is optimized again to maximize $\overline{\chi}$ under the conditions of the experimental setup. The blue line shows the numerical CD spectrum of the new design in Fig.~\ref{fig:experimental_results}a. The chirality of this new design reaches a value of 0.76 at the design wavelength. The parameters of the optimal helix are indicated in the caption of Fig.~\ref{fig:SEM}. The orange line of Fig.~\ref{fig:experimental_results} shows the CD signal from the experimental measurement. One can observe how both CD spectra differ considerably in magnitude and dispersion. 

To find a reason for the discrepancy, we look at SEM images of the fabricated helices before their release from the substrate into solution in Fig.~\ref{fig:SEM}a. From the image one can notice that the fabricated helices differ from the optimal design, Fig.~\ref{fig:SEM}b, mainly in the pitch. Also, some of the fabricated helices appear to be slightly tilted and feature a tapered shape. We stress that the fabrication is challenging. To test if the difference in geometry was the source of the discrepancy between the predicted and obtained measurements, we generated a CAD model of a helix that better resembles the shapes appearing in the SEM image. The shape of this revised CAD model is shown in Fig.~\ref{fig:SEM}c. However, the geometry adjustment did not provide a much better agreement between the experimental and numerical results. 

This improved agreement appeared when we took into consideration a likely oxidation process of the helices. To check this hypothesis, we calculated new values for the silver permittivity of the helix as a weighted sum with a 70 percent of silver\cite{mcpeak2015plasmonic} and a 30 percent of partially oxidated silver. The permittivity values of the silver oxide were extracted from Fig.~1 of Ref.~\mbox{(\!\!\citenum{qiu2005ellipsometric})}, using the plot line for the lower flux of oxygen. The new numerical results obtained, also using the CAD model of Fig.~\ref{fig:SEM}c, are shown in Fig.~\ref{fig:experimental_results}b. One can see a much better agreement between the experimental and these newly computed results. Nevertheless, one can still appreciate a wavelength shift between both results. We assume that this difference might be due to the material properties of the helices, because reasonable modifications of the geometry parameters of the CAD model, in agreement with the observed differences in the SEM image, did not shift the peaks of the resonances significantly. The experimental results indicate that the fabrication of optimal helices at optical frequencies is a fabrication challenge which may be tackled for the fabrication technique used here, and also for other techniques \cite{Hoeflich2011,Wozniak2018}.


\section{Omni-directional filter}

To give an estimate of the implications of different values of the em-chirality $\overline{\chi}$ of a scattering object, we now look at the performance of some of the analyzed helices when used to create an omni-angle polarization filter. In this comparison, we consider the optimal helix obtained for a design wavelength of $\lambda _{\mathrm{d}}$=3~\micras, the optimal helix obtained for a design wavelength of $\lambda _{\mathrm{d}}$ = 800~nm and embedded in water, and the tapered helix obtained as a result of trying to match the experimental and the numerical results. 
To make this comparison, we use a rather simple approximation of the filter. We will assume that the filter, composed of randomly oriented helices embedded in a homogeneous medium with the refractive index of water, can be treated as a single homogenized material. We consider that the number of embedded helices is sufficiently large that the total resultant scattering effects can be neglected. The absorption that this material presents to light of different helicity can be approximated using the rotationally averaged helicity absorption cross-sections\cite{fernandez2020computation,fazel2021orientation} of the helices, $\sigma _{\mathrm{abs},\pm}$. 
If we consider a slab of this homogenized material with a thickness $\Delta z$, assuming that it is thin enough for the power flux of the incident beam remaining approximately constant, the power per unit area absorbed by the medium with a helix number density $\rho_{\mathrm{helices}}$ can be obtained as,

\begin{equation}
    S_{z}(z+\Delta z)-S_{z}(z) = -\Delta z \rho_{\mathrm{helices}} S_{z}(z)\sigma _{\mathrm{abs},\pm},
\end{equation}
with $S_z$ being the component of the time averaged Poynting vector along the propagation direction of the beam.
In the limit $\Delta z \to 0$, one obtains that the change of the power flux along the propagation direction of the beam equals,
\begin{equation}
\frac{\mathrm{d}S_z(z)}{\mathrm{d}z} = -S_z (z)\, \rho _{\mathrm{helices}}\,\sigma_{\mathrm{abs},\pm},
\end{equation}
which gives us the value of the power flux of the beam as a function of the initial power flux, $S_{z,0}$, before entering the slab of homogenized material,
\begin{equation}\label{eq:absorption_helices}
    S_z(z) = S_{z,0}e^{-\rho _{\mathrm{helices}}\sigma_{\mathrm{abs},\pm}z},
\end{equation}

\begin{figure}[h!]
\centering\includegraphics[width=0.95\linewidth]{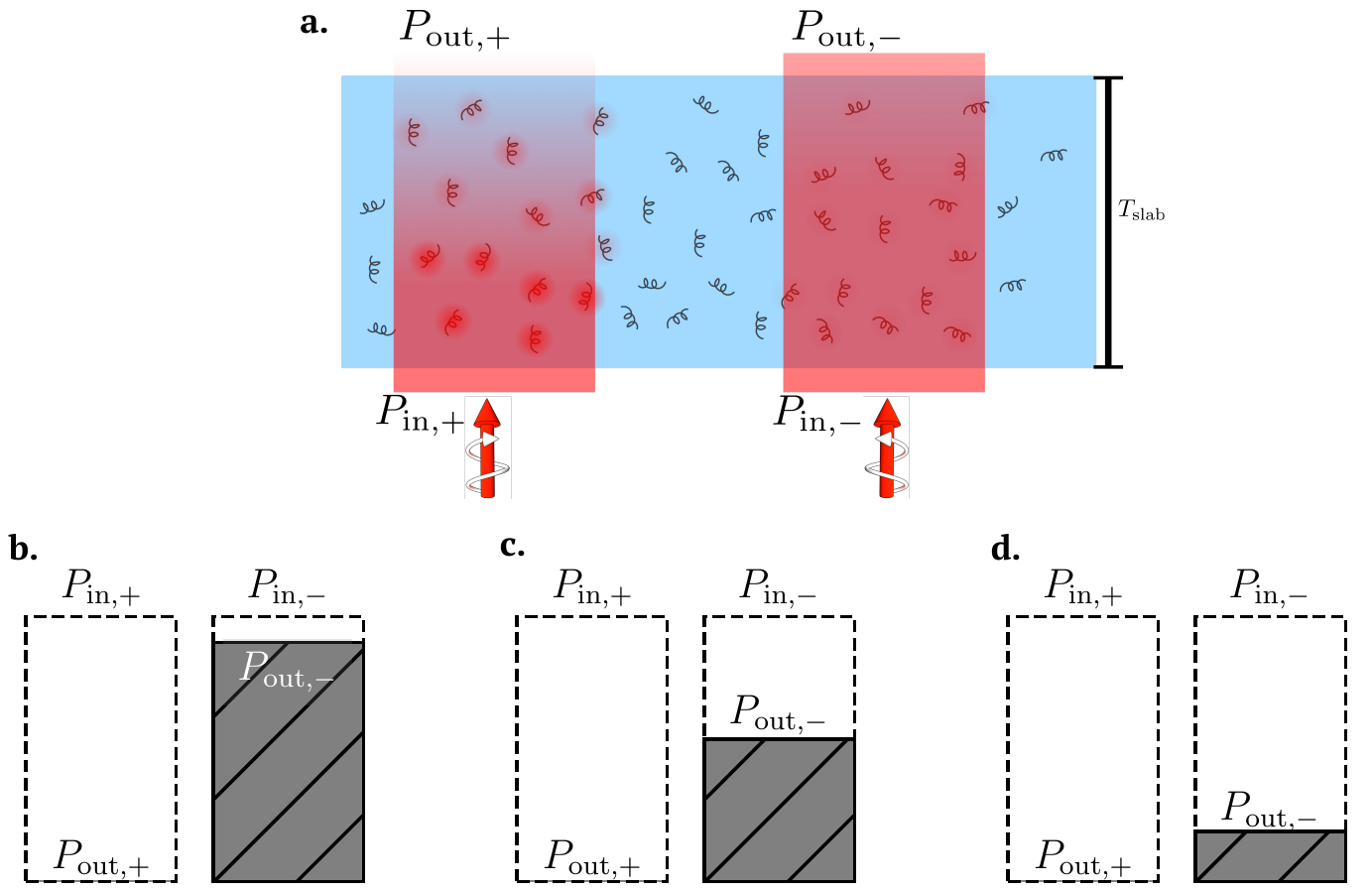}
\caption{\textbf{a}: Sketch of the omni-directional polarization filter concept. The filter consist of a embedding medium containing randomly positioned and oriented silver helices with the same winding direction. \textbf{b},\textbf{c} and \textbf{d}: Performance of three different omni-directional polarization filters. The bars show which fraction of the input power is transmitted to the other side of the filter when illuminated with beams of different helicity. \textbf{a}: Filter made out of helices with the design obtained from the optimization at $\lambda _{\mathrm{d}}$ = 3~\micras{}. The performance of the filter is measured at a vacuum wavelength $\lambda _0$ of 3~\micras{}. \textbf{b}: Filter made out of helices obtained from the optimization result at the design wavelength $\lambda _{\mathrm{d}}$ = 800~nm. The performance of the filter is measured at a vacuum wavelength $\lambda _0$ of 800 nm. \textbf{c}: Filter made out of the silver oxidized helices model shown in the previous section. The performance of the filter is measured at a vacuum wavelength $\lambda _0$ of 740 nm. From left to right, the filters have thicknesses of 7.7~\micras{}, 21.4~\micras{} and 19.8~\micras{} respectively.}
\label{fig:polarization_filter}
\end{figure} 

To compare how the filters would absorb beams of different helicity, we set the thickness of the homogenized slab such that it absorbs 99 percent of the incoming power for the beam with helicity +1. We will use a helix volume filling fraction of 5 percent. This quantity turns out to be a slab of 7.7~\micras{} for the material composed of optimal helices for the design wavelength of 3~\micras{}, 21.4~\micras{} for the material composed of  optimal helices obtained for the design wavelength of 800~nm, and 19.8~\micras{} for a material composed of the tapered silver oxidized helices. 
The results for the performance of the three filters are shown in Fig.~\ref{fig:polarization_filter}. The figure shows the transmittance of the filters for illumination beams of both helicities. For the case of the incoming beams with helicity -1, the slab with the helices designed at $\lambda _{\mathrm{d}}$ = 3~\micras{} lets 91 percent of the incoming power flux through, the slab with helices designed at $\lambda _{\mathrm{d}}$ = 800~nm transmits 54 percent, and the slab composed of the tapered silver oxidized helices transmits 19 percent of the incoming power flux. 

From right to left, Fig.~\ref{fig:polarization_filter} shows the performance of a device that can be currently fabricated, one device that could be obtained if one could further improve the fabrication process, and one that, for achieving a similar behavior at optical frequencies, further theoretical work needs to be done to obtain a suitable design.

\section{Conclusion}

We have optimized silver helices for maximizing their electromagnetic chirality at particular frequencies ranging from the far infrared to the optical band. Values larger than 90 percent of the maximum possible value have been obtained from the lowest frequency up to a wavelength of \SI{3}{\micro\meter}. As the frequency increases into the optical band, the electromagnetic chirality of the helices decreases monotonically to just below 0.6 at \SI{500}{\nano\meter}.

For the optimization, we have combined a method for computing shape derivatives for T-matrices of complex structures with a Bayesian optimization algorithm that supports the use of derivative information.

In accordance with their high electromagnetic chirality, the optimized helices show a strong difference in the interaction with fields of different handedness, including the rotationally averaged differential interaction and absorption cross-sections. These differences make them suitable components in devices for different applications such as helicity filtering glasses. We demonstrate that, at the operational wavelength of \SI{3}{\micro\meter}, a \SI{8}{\micro\meter} slab of randomly arranged helices at 5 percent filling factor is able to absorb 99\% of the light of unwanted helicity while absorbing only 9 percent of the light of the desired helicity.

Further theoretical research is necessary to find different materials and designs that enable a similar behavior deep in the optical regime and possibly across an extended spectral domain. A basis for the application of a wide variety of mathematical optimization techniques has been provided by integrating em-chirality with mathematical formulations of scattering theory\cite{Arens2018}. This in principle opens up the possibility of applying general free-form shape optimization techniques to the design problem. A first, but already very promising option is to extend the shapes of the design space from helices to more complex free-form wires\cite{Capdeboscq2021,Arens2021}.

Moreover, our experimental results for a design at \SI{800}{\nano\meter} show that the performance of these structures strongly depends on an accurate realization of their optimal material properties, and optimal shapes, with features as small as a few tens of nanometers.

Unlocking the applications of high em-chiral objects at optical frequencies hence implies addressing both theoretical and fabrication challenges.

\section{Methods}\label{sec:Methods}

\subsection{T-matrix calculation}

The T-matrices of the silver helices are computed using the finite element solver JCMsuite\cite{pomplun2007adaptive,burger2008jcmsuite}. They are computed by illuminating the helix with a set of regular vector spherical wave functions, and the scattered fields produced by each illumination are expanded into a set of radiative vector spherical wave functions. The procedure is the same as described in Ref.~\mbox{(\!\!\citenum{demesy2018scattering})}. The only difference lies in the numerical method we use to expand the scattered fields, and which in our case is the one that we previously presented in Ref.~\mbox{(\!\!\citenum{garcia_santiago2019decomposition})}.
We use finite elements with a polynomial degree of 2. We need small mesh elements to reproduce the surface of the helix accurately and therefore do not benefit from higher polynomial degrees. The accuracy requirements would already be fulfilled because of the small mesh size. The maximum mesh size used for the helix domain is $\mathrm{\lambda}$/0.3 for the longer wavelengths and $\mathrm{\lambda}/2$ for the optical and near-infrared wavelengths. Note that $\mathrm{\lambda}$ refers to the wavelength within the helix, and the refractive index of silver at the longer wavelengths exceeds values of 300 in both the real and imaginary parts.
In the air domain, we use a mesh size of $\mathrm{\lambda}$/15. Before starting a new optimization at each wavelength, a convergence test is done to know if further mesh refinement is needed. The criterion is to have a fine mesh to achieve errors in $\overline{\chi}$ lower than one percent.

\subsection{Calculation of the derivatives of the T-matrices concerning the geometrical parameters of the helix}

The shape derivatives are computed in a multi-step process. Once a tetrahedralization of the helix is computed, a separate python script is used to read the mesh points that lie on the helix surface and subsequently obtain the parameters of the analytic parametrization of the helix corresponding to those points. Once the set of parameters is obtained, the derivatives are computed based on analytic expressions. They are passed back to JCMsuite, which finalizes the process of propagating the shape derivatives to the calculation of the T-matrix using the forward method\cite{kepler2010sensitivity,burger2013fast,hughes2019forward}.

\subsection{Optimization process}

The optimization of the helices consisted of different optimization runs, each for the optimization of a helix at a different design wavelength, $\lambda_{\mathrm{d}}$. 

The optimization algorithm used in each optimization run was Bayesian optimization, more specifically, we used the optimization tool included into the software JCMsuite. After each new FEM simulation was computed, we included the value of the calculated $\overline{\chi}$ plus its derivatives with respect to the four design parameters: $R_{\mathrm{h}}$, $T_{\mathrm{w}}$, $P_{\mathrm{h}}$, and $N_{\mathrm{turns}}$ into the Gaussian process of the optimization algorithm.

The limits of the parameter space for the different optimization runs were set downscaling the ones used in the first optimization run for $\lambda_{\mathrm{d}}$ = 150~\micras{}. At the illumination wavelength of 150~\micras{}, the lower and upper limits for the radius of the helix, $R_{\mathrm{h}}$, were set to 1 and 10~\micras{}, respectively. When we decrease the operational wavelength, they were accordingly linearly scaled down. The thickness of the wire, $T_{\mathrm{w}}$, was allowed to vary between 5 percent and 95 percent of the value of the radius of the helix. An additional lower limit was imposed on the thickness of the wire. It could not be smaller than 20~nm for any of the simulations, independently of the wavelength. This reflected fabrication limits in making extremely thin wires. It also accommodated the lower bound for the critical dimension of the radius of the wire to rely on macroscopic Maxwell's equations equipped with bulk constitutive relations to describe the light-matter-interaction~(e.g., Ref.~\mbox{(\!\!\citenum{jackson1999classical})} section 6.6). The pitch of the helix, $P_{\mathrm{h}}$, was set relative to the thickness of the wire. It could vary between 1.1 and 10 times the thickness. The limits for the number of turns were 0.1 and 4.

Each optimization run consisted of 500 finite element simulations of different helices of the design space. The parameters to be simulated were chosen by the Bayesian optimization algorithm. After the 500 simulations, the optimization process would continue if the value of the expected improvement\cite{jones2001taxonomy,hennig2012entropy}, obtained from the Gaussian process, would be higher than $10^{-12}$. However, this was not the case for any of the optimization runs and all of them stopped after the 500 iterations. 
Regarding the optimization time, the time required to evaluate the em-chirality, $\overline{\chi}$, at each point of the design space depended on the specific design wavelength and specific size of each helix. On average, each finite element simulation took around 6 minutes.

\subsection{Experimental realization of the helices}

Dedicated helices were fabricated by a published shadow growth physical vapor deposition technique that combines glancing angle deposition (GLAD) with nanopatterning (nanoGLAD)\cite{mark2013hybrid}. Briefly, a silicon wafer was cleaned and then pre-patterned with a monolayer of hexagonally arranged gold nanoparticles  using  Block-Copolymer-Micelle-Lithography  (BCML)\cite{glass2003block}. Silver(Ag) and copper(Cu) were then heated by an electron beam and deposited onto the patterned wafer under high vacuum. The  angle between the flux direction of the incident atoms and the substrate normal was $\alpha$=87º. The gold nanoparticles facilitate shadowing and act as seeds for the subsequent growth. The spacing between the seeds was $d$= 76 nm. During the deposition process, the substrate was cooled with liquid nitrogen and slowly rotated about the substrate normal, leading to the formation of helically shaped $\mathrm{Ag_{0.95}Cu_{0.05}}$ nanostructures with 2.4 turns. After the fabrication was completed,  the particles were removed from the substrate and transferred into (Milli-Q) water via ultrasonication. The geometric helix dimensions were extracted from scanning electron micrographs (pitch $P_{\mathrm{h}}= 72.5\pm3.5$ nm, wire width $T_{\mathrm{w}} = 44.1\pm6.6$ nm, radius $R_{\mathrm{h}}= 23.5\pm 6.7$ nm), an example image of the fabricated helices is presented in Fig.\ref{fig:SEM}a. The sonicated wafer piece had an area of $A_\mathrm{w}~\mathrm{\approx 25 \cdot 10^{-6}}$~$\si{\metre}^2$. We thus estimate that the colloidal suspension contains 
\begin{equation}
N_\mathrm{p}= \frac{2\cdot A_\mathrm{w}}{\sqrt{3}\cdot d^2} = 5\cdot 10^9
\end{equation}
particles. The  particles  have  been  suspended in 1.7 ml of water. Hence, the number density of the colloidal solution containing the helices was $N= 2.9\cdot 10^{15}\,\si{\metre}^{-3}$. However, it is important to realize that this value only reflects an upper bound.  In practice, imperfections in the seed layer or during the fabrication result in fewer particles on the wafer. We thus estimate a number density for the colloidal solution of $N=\left(1.6\pm 1.3\right)\cdot 10^{15}\,\si{\metre}^{-3}$.


\begin{acknowledgement}
The authors gratefully acknowledge financial support by the Deutsche Forschungsgemeinschaft (DFG, German Research Foundation) through Project-ID 258734477 - SFB 1173 and Project-ID 390761711 - EXC 2082/1, from the European Union's Horizon 2020 research and innovation programme under the Marie Sklodowska-Curie grant agreement No 675745, by the Helmholtz Association via the Helmholtz program “Materials Systems Engineering” (MSE), and the KIT through the “Virtual Materials Design” (VIRTMAT) project.  We acknowledge support from the Karlsruhe School of Optics and Photonics (KSOP) and from the Carl Zeiss Foundation via the CZF-Focus@HEiKA program. This project has received funding from the German Federal Ministry of Education and Research (BMBF Forschungscampus  MODAL, project number 05M20ZBM) and from the Deutsche Forschungsgemeinschaft (Excellence center MATH+, EXC-2046/1, project ID: 390685689).
\end{acknowledgement}


\bibliography{bibliography_paper_em_chiral_helices}

\end{document}